\documentclass[prl,preprint,superscriptaddress]{revtex4-2}
\usepackage{dcolumn}
\usepackage{mathtools}
\usepackage{amssymb}
\usepackage{bm}
\usepackage{pifont}
\usepackage{psfrag}
\usepackage{epstopdf}
\usepackage{wasysym}
\usepackage{amsmath}
\usepackage{tipa}
\usepackage{hyperref}
\usepackage{verbatim}
\usepackage{graphicx}
\usepackage{epsfig}
\usepackage[usenames]{color}
\usepackage{xcolor}

\begin{document}

\title{Motile topological defects hinder dynamical arrest in dense liquids of active ellipsoids}
\author{Pragya Arora}
\affiliation{Chemistry and Physics of Materials Unit, Jawaharlal Nehru Centre for Advanced Scientific Research, Jakkur, Bangalore - 560064, INDIA}
\author{A K Sood}
\affiliation{Department of Physics, Indian Institute of Science, Bangalore- 560012, INDIA}
\affiliation{International Centre for Materials Science, Jawaharlal Nehru Centre for Advanced Scientific Research, Jakkur, Bangalore - 560064, INDIA}
\author{Rajesh Ganapathy}
\affiliation{International Centre for Materials Science, Jawaharlal Nehru Centre for Advanced Scientific Research, Jakkur, Bangalore - 560064, INDIA}
\affiliation{School of Advanced Materials (SAMat), Jawaharlal Nehru Centre for Advanced Scientific Research, Jakkur, Bangalore - 560064, INDIA}
\date{\today}

\begin{abstract}
\textbf{Recent numerical studies have identified the persistence time of active motion as a critical parameter governing glassy dynamics in dense active matter. Here we studied dynamics in liquids of granular active ellipsoids with tunable persistence and velocity. We show that increasing the persistence time at moderate supercooling is equivalent to increasing the strength of attraction in equilibrium liquids and results in reentrant dynamics not just in the translational degrees of freedom, as anticipated, but also in the orientational ones. However, at high densities, motile topological defects, unique to active liquids of elongated particles,  hindered dynamical arrest. Most remarkably, for the highest activity, we observed intermittent dynamics due to the jamming-unjamming of these defects for the first time. }
\end{abstract}

\maketitle

Physical intuition would suggest that making particles' in a supercooled liquid active, i.e., self-propelled, should shift the glass transition to higher densities/lower temperatures. Determining if this is indeed the case has been the mainstay of active glass research since experiments found that dynamics in confluent epithelial cell sheets shared remarkable similarities with equilibrium supercooled liquids and glasses \cite{angelini2011glass}. The many subsequent numerical \cite{berthier2014nonequilibrium, levis2014clustering, ni2013pushing, mandal2017glassy, flenner2016nonequilibrium} and analytical studies \cite{szamel2015glassy, szamel2016theory, feng2017mode} now anticipate that how the glass transition shifts depends explicitly on how the active forces are introduced \cite{nandi2018random}, the form of the particle interactions, and crucially, the persistence time of particles' directed motion, $\tau_p$ \cite{berthier2017active}. In one class of well-studied active glass models, where the particle dynamics follow an Ornstein-Uhlenbeck stochastic process, the glass transition in active liquids of hard spheres shifts to higher densities with increasing $\tau_p$ \cite{berthier2014nonequilibrium}. For softer interactions, surprisingly, this shift is towards higher temperatures/lower densities because a large $\tau_p$ manifests as an effective swelling of the particles \cite{flenner2016nonequilibrium}. Since a finite persistence results in an activity-mediated effective attraction between even hard particles and leads to clustering \cite{levis2014clustering, cates2015motility}, this model also anticipates re-entrant liquid dynamics with $\tau_p$ for intermediate values of particle softness and even a re-entrant glass transition \cite{berthier2017active}.  This suggests a possible mapping with equilibrium `sticky' hard particle liquids where such a re-entrant behavior is observed with increasing attraction strength \cite{pham2002multiple, eckert2002re, mishra2013two}. Neither the increased tendency to glassify nor the re-entrant dynamics have been seen in experiments largely because tuning $\tau_p$ systematically has not been possible hitherto. Importantly, even while model details differ, studies anticipate that there is always a glass transition \cite{berthier2013non, liluashvili2017mode} and dynamical slowing down is accompanied by large collective swirls \cite{berthier2014nonequilibrium, levis2014clustering, ni2013pushing, mandal2017glassy, flenner2016nonequilibrium, szamel2015glassy, szamel2016theory, feng2017mode}. These swirls have indeed been observed in living active matter experiments
\cite{angelini2011glass, garcia2015physics, henkes2020dense}.

Even though numerical/theoretical active glass studies take inspiration from findings in biological systems, where the constituents are anisotropic in shape, the models, barring the one on dumbbells \cite{mandal2017glassy}, are for isotropic particle assemblies \cite{berthier2017active}. When alignment interactions are present, swirls are more pronounced \cite{mandal2017glassy}, and furthermore, shape anisotropy invariably results in topological defects, which in active matter can be motile \cite{doostmohammadi2018active}. These defects play a crucial role in tissue morphogenesis, regeneration, and migration \cite{saw2017topological, kawaguchi2017topological, hirst2017liquid} although their role in glassy slowing down remains unexplored.

Here, we developed a system of active granular ellipsoids where $\tau_p$ and the particle velocity, $v$, can be tuned systematically while keeping the particle shape features fixed. Just like in the simulations of active Ornstein-Uhlenbeck particles (AOUPs) \cite{feng2017mode}, the thermal noise is irrelevant here. We show that at mild to moderate supercooling the evolution of dynamics on increasing activity is akin to a system of hot adhesive particles with re-entrant dynamics in both the translational and orientational DOF. A mode-coupling theory (MCT) inspired analysis reveals an increased tendency to glassify at the largest $\tau_p$. Near the MCT glass transition and beyond, motile topological defects acted as conduits for structural relaxation and suppressed glassy dynamics. Remarkably, at the largest $\tau_p$, we observed intermittent jamming-unjamming events that are strikingly similar to that predicted to occur in so-called `extreme' active matter - systems where $\tau_p$ is very large \cite{mandal2020extreme}.

Our experimental system comprised of 3D-printed prolate  ellipsoids (major-axis $\alpha = 6$ mm and minor-axes $\beta = 2.4$ mm, $\delta = 2.1$ mm) rendered active by vertical vibration. Earlier studies found that granules with an asymmetry in shape or mass, or friction coefficient, $\mu$, or some combination of these, under vertical agitation were polar active along the direction set by the asymmetry
 \cite{kudrolli2008swarming,deseigne2010collective,
 kumar2014flocking, arora2021emergent}.   To tune activity, i.e. $\tau_p$ and/or $v$, the particle shape or size cannot be altered since this itself can strongly influence the nature of glassy dynamics. Instead, we designed particles with a fore-aft asymmetry in $\mu$ as well as mass. The former was brought about by exploiting a feature unique to the print process and the latter by introducing a hole along the particles' major-axis (Supplemental Material S1, Fig. S1 and Fig. S2) \cite{arora2021emergent}. Since changing the hole positions alters the particles' center-of-mass, we expected this to allow tuning activity.  

\begin{figure}[tbp]
\includegraphics[width=0.75\textwidth]{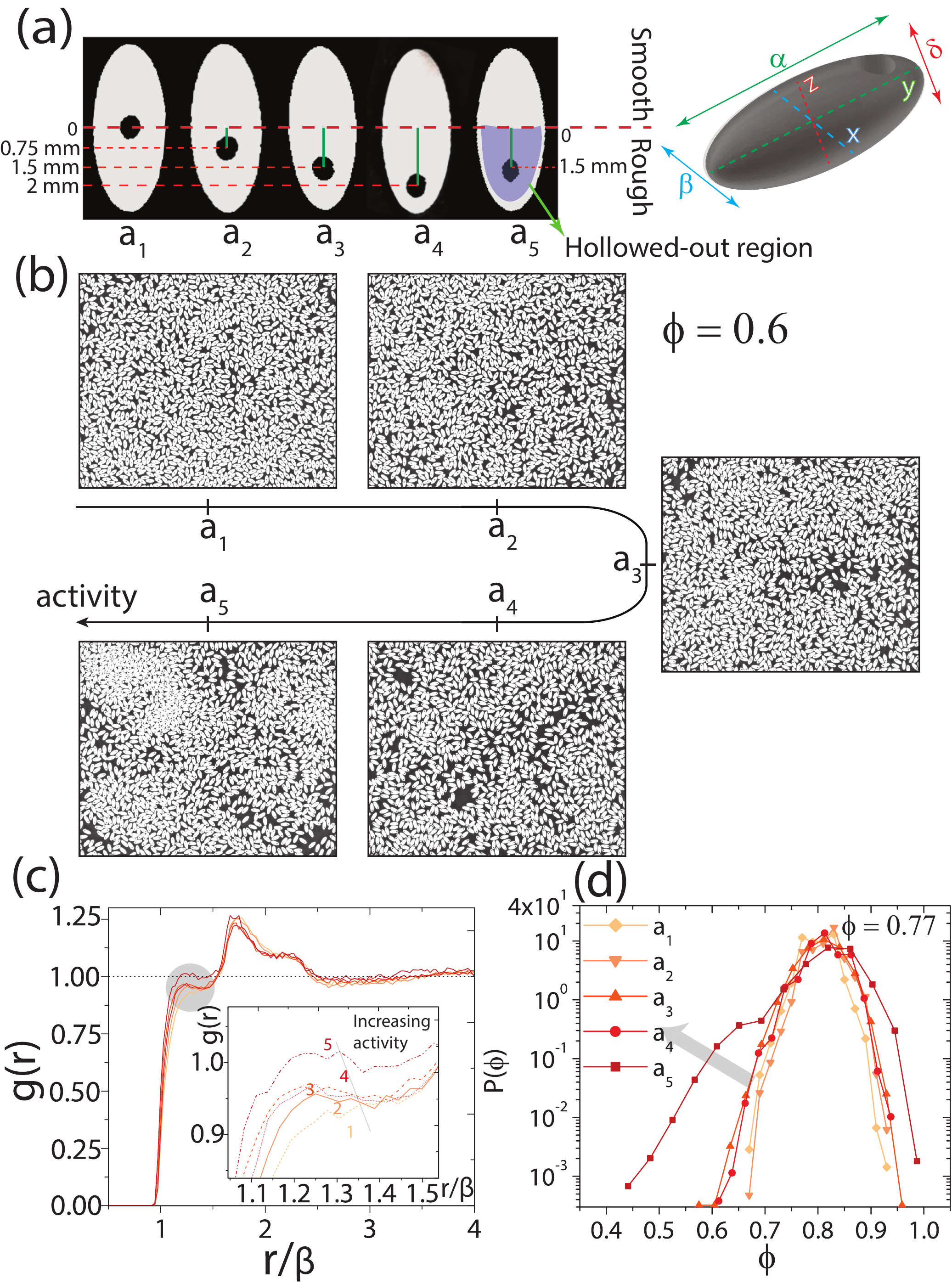}
\caption{(a) Left panel: Snapshots of the 3D-printed active ellipsoids.  The central red dashed lines divides the rough portion of the particle from the smooth side. The hole position is progressively shifted across the major-axis for $a_{1}$ to $a_{4}$. For $a_{5}$, one half of the ellipsoid is hollowed out and is shown by the blue shaded region. Right panel: 3D view of the ellipsoid showing its dimensions. (b) Snapshots of the experiment for $\phi=0.6$ as a function of activity.  Increasing the activity induces a strong tendency for particles to cluster. (c) Radial pair-correlation function,  $g(r)$ at various activities. The inset shows the zoomed in region corresponding to the first peak of the $g(r)$. (d) The probability distribution of the coarse-grained density field, ${P}(\phi)$, at various activities.
}
\label{Figure1}
\end{figure}

Figure \ref{Figure1}(a) shows snapshots of the ellipsoids for various hole positions and labelled $a_1$ through $a_5$. To further enhance the mass asymmetry, for $a_5$, one half of the ellipsoid was also made hollow during print (shaded region). The particles were placed on a horizontal plate that was coupled to an electromagnetic shaker through an air bearing and confined from above by a glass plate to enable imaging of dynamics \cite{harris2015generating}. The gap between plates $\Delta$ is such that $\delta<\Delta<\beta$ and particles cannot flip once confined. The drive frequency $f = 37$ Hz and amplitude $a  = 1$ mm were kept constant, and the non-dimensional acceleration $\Gamma = {4\pi^2f^2a\over g} = 5.5$, where $g$ is the acceleration due to gravity. For all activities, the particle motion was typical of run-and-tumble dynamics ( Movie S1), and as anticipated, both $\tau_p$ and $v$ increased largely systematically (Supplemental Material S2,  Fig. S3 and Table. S1).

Armed with the capability to tune activity without changing the particle shape/size, we investigated the behaviour of dense assemblies. For each activity, the projected particle area fraction, $\phi$, was varied from $0.54$ to $\sim 0.84$. With increasing activity, we evidenced a clear and concomitant increase in the tendency for particles to form clusters (Fig. \ref{Figure1}(b) for $\phi=0.6$  and  Movie S2 and S3 \cite{levis2014clustering, cates2015motility,fily2014freezing}. This is quantitatively captured by the radial pair-correlation function, $g(r)$, where the position of the first peak, which corresponds to the lateral alignment of the ellipsoids, remains fixed, while its height increases systematically with activity due to a higher likelihood of finding a nearby neighbor (Fig. \ref{Figure1}(c) and inset). Clearly, increasing activity has a  role similar to increasing the strength of short-range attraction in equilibrium systems \cite{berthier2014nonequilibrium, levis2014clustering, berthier2017active}. However, the clustering we observed was only transient indicating that this may not be motility-induced phase separation (MIPS) which manifests as a stable, dense macro droplet in a dilute gas-like background and is seen as a clear bi-modal in the probability distribution of the coarse-grained density field, $P(\phi)$ (see Movie S4) \cite{cates2015motility}. While with increase activity, $P(\phi)$ did develop a shoulder (Fig. \ref{Figure1}(d)), its height is almost an order-of-magnitude smaller than the peak value of $P(\phi)$ (See Supplemental Material S4 and   Fig. S4, Fig. S5  for other degrees of coarse graining). Thus, unlike in isotropic active particle assemblies, MIPS is relatively weak here. Although surface roughness, like in our ellipsoids, can promote MIPS as it increases the particle collision time \cite{ilse2016surface}, our system did not show large-scale phase separation possibly due to the counter-effect of the rod-like excluded volume torque \cite{van2019interparticle}. 

\begin{figure}[tbp]
\includegraphics[width=0.9\textwidth]{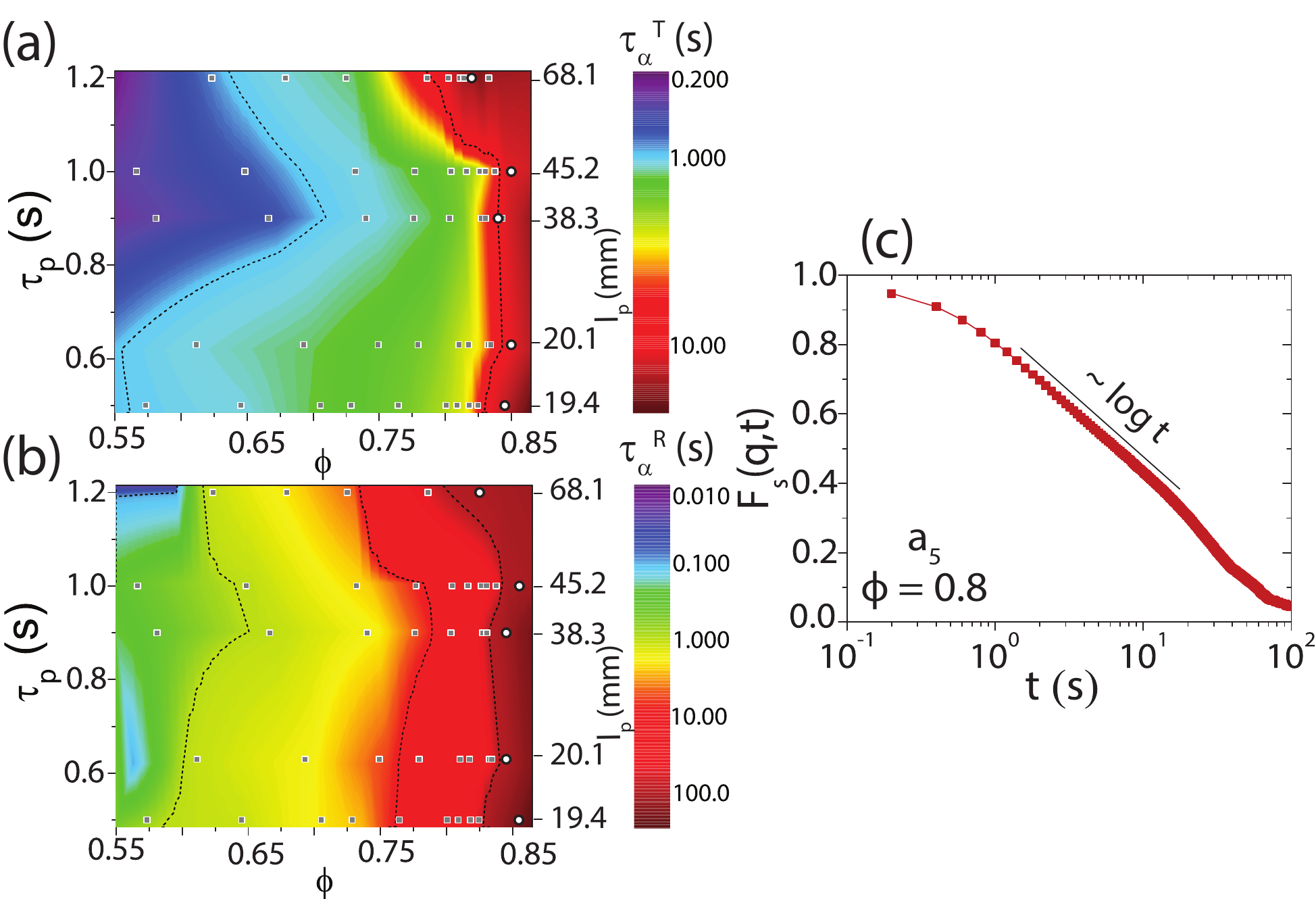}
\caption{(a and b) show the relaxation dynamics phase diagram in the ($\tau_{p}$ , $\phi$) plane for the translational and orientational degrees of freedom (DOF), respectively. The squares represent the $\tau_{p}$ and $\phi$ at which experiments were performed. In (a) and (b) the isochrones  are shown by black dashed line. The color bar indicates the value of  $\tau_{\alpha}^T$ and $\tau_{\alpha}^R$ . The $\tau_{\alpha}$ values for $\phi$'s in between experimental data points were obtained from a linear interpolation. The second label on y-axis indicates the persistence length $l_{p}$. $\phi_{g}^{T}$ and $\phi_{g}^{R}$, are shown by circles in (a) and (b) respectively. (c) $F_{s}(q=1.4 \text{mm}^{-1},t)$ versus time for $\phi=0.8$ for $a_{5}$.
} 
\label{Figure2}
\end{figure}

Next, we interrogated how the activity-mediated effective attraction influenced dynamical slowing down. We quantified the relaxation dynamics of the system through the translational and orientational structural relaxation times $\tau_\alpha^T$ and $\tau_\alpha^R$, respectively,  obtained from the corresponding time correlators, these being the self-intermediate scattering function, $F_{s}(q, t)$, and the $n^{\text{th}}$-order orientation correlation function, $L_{n}(t)$(Supplemental Material S5 and Fig. S6). Figure \ref{Figure2}(a) and (b) show the phase diagram of active ellipsoids in the ($\tau_{p}$, $\phi$)  plane with the background colours representing  the magnitude of $\tau_{\alpha}^R$ and $\tau_{\alpha}^T$ on a logarithmic scale. The second label on y-axis indicates the persistence length  $l_{p}$ = $v$ $\tau_{p}$ (see Fig. S3 and Table S1). The gray squares represent the values of $\tau_{p}$ and $\phi$ at which experiments were performed and the isochrones are shown as black dashed lines. Consider, a vertical cut at $\phi = 0.65$, increasing activity first speeds up relaxation and is then followed by a slowing down at the highest activities. This re-entrant dynamics is a consequence of the competition between particle crowding and self-propulsion\cite{berthier2014nonequilibrium, levis2014clustering, berthier2017active}. Initially, increasing $\tau_p$ leads to clustering of ellipsoids which opens up voids where self-propulsion dominates resulting in a speeding up of dynamics. However, on further increasing  activity (effective attraction), clustering dominates and slows down dynamics. This constitutes the first observation of re-entrant dynamics in the liquid phase of active elongated particles.

A recently developed MCT for isotropic AOUPs, predicts a non-monotonic evolution of $\tau_{\alpha}^T$ with $\tau_p$ \cite{szamel2015glassy, feng2017mode}. Furthermore, a similar evolution of the glass transition with $\tau_p$ is also expected in a small region of the parameter space \cite{berthier2017active}. There are, however, no such predictions for dense liquids of active ellipsoids as yet. Motivated by these studies, we fitted the relaxation times in the mild to moderate supercooled regime to power-laws of the form, $\tau_{\alpha}^{R,T} \propto \left(\phi_g^{R,T} / \phi-1\right)^{-\gamma}$, and determined the orientational and translational glass transition area fractions, $\phi_g^R$ and $\phi_g^T$, respectively.  Here, $\gamma$ is the scaling exponent (See Fig. S7 and Fig. S8). The black circles in Fig. \ref{Figure2}(a) and (b) represent $\phi_{g}^T$ and $\phi_{g}^R$, respectively. Although $\phi_{g}^T$ and $\phi_{g}^R$ shift to a smaller $\phi$ at the highest activity, we did not observe the stark re-entrant behaviour observed at moderate supercooling. Nevertheless, for  $a_{5}$, $F_{s}(q, t)$ at $\phi = 0.8$ showed a logarithmic decay over more than a decade in $t$ (Fig. \ref{Figure2}(c)). MCT first predicted such an anomalously slow relaxation for equilibrium `sticky' hard spheres near the $A_3$ singularity - the point where the attractive and repulsive glass transition lines meet \cite{dawson2000higher} - and was later seen in many different passive liquids \cite{chen2003glass, pham2004glasses, mishra2013two} but never in active ones. These observations suggest that the change in dynamics brought about by increasing activity mirrors the change due to an increase in the strength of inter-particle attraction in equilibrium systems.

The above analysis revealed another correspondence between dense active and passive liquids of ellipsoids. For our active ellipsoids, which have an aspect ratio $p={\alpha\over\beta} = 2.5$, the $\phi_g^R$ and $\phi_g^T$ values are nearly same for each individual activity (Fig. S9). MCT for passive liquids of hard ellipsoids predicts orientational and translational glass transitions to occur at the same $\phi$ when $p\leq 2.5$, and orientational freezing to precede the translational one for $p>2.5$ \cite{mishra2013two, letz2000ideal, zheng2011glass}. This correspondence with passive liquids is intriguing since a key ingredient of Ref. \cite{flenner2016nonequilibrium} is equal-time spatial velocity correlations, which should promote alignment interactions and aid slow down dynamics in the orientational DOF.

A close inspection of the particle dynamics did reveal swirls at  moderate supercooling, and importantly, their macroscopic ordering was nematic, although our ellipsoids are polar active (Supplemental Material S8 and Fig. S10) \cite{doostmohammadi2018active, grossmann2020particle}. We extracted the correlation length, $\zeta$, of the swirls from the angle-averaged correlation function of  the orientation of the displacement vectors of the active ellipsoids, $C(r)=\left\langle 2 \cos ^{2} \Delta \theta(\mathbf{r})-1\right\rangle$ ( See Supplemental Material S9 and Fig. S11) \cite{mandal2017glassy}. Here, $\Delta \theta(\mathbf{r})$ is the angular separation between two displacement vectors separated by distance $r$, and the displacement was calculated over $\tau_{\alpha}^T$ since our goal was only to determine how these swirls influenced structural relaxation. For $\phi<0.78$, the regime that largely determines the $\phi_g$ values, and at all activities, $\zeta$ remains small (Fig. S12). Thus, our active liquids behave like passive ones, and hence also the striking agreement with MCT predictions for the latter.   

\begin{figure}[tbp]
\includegraphics[width=0.9\textwidth]{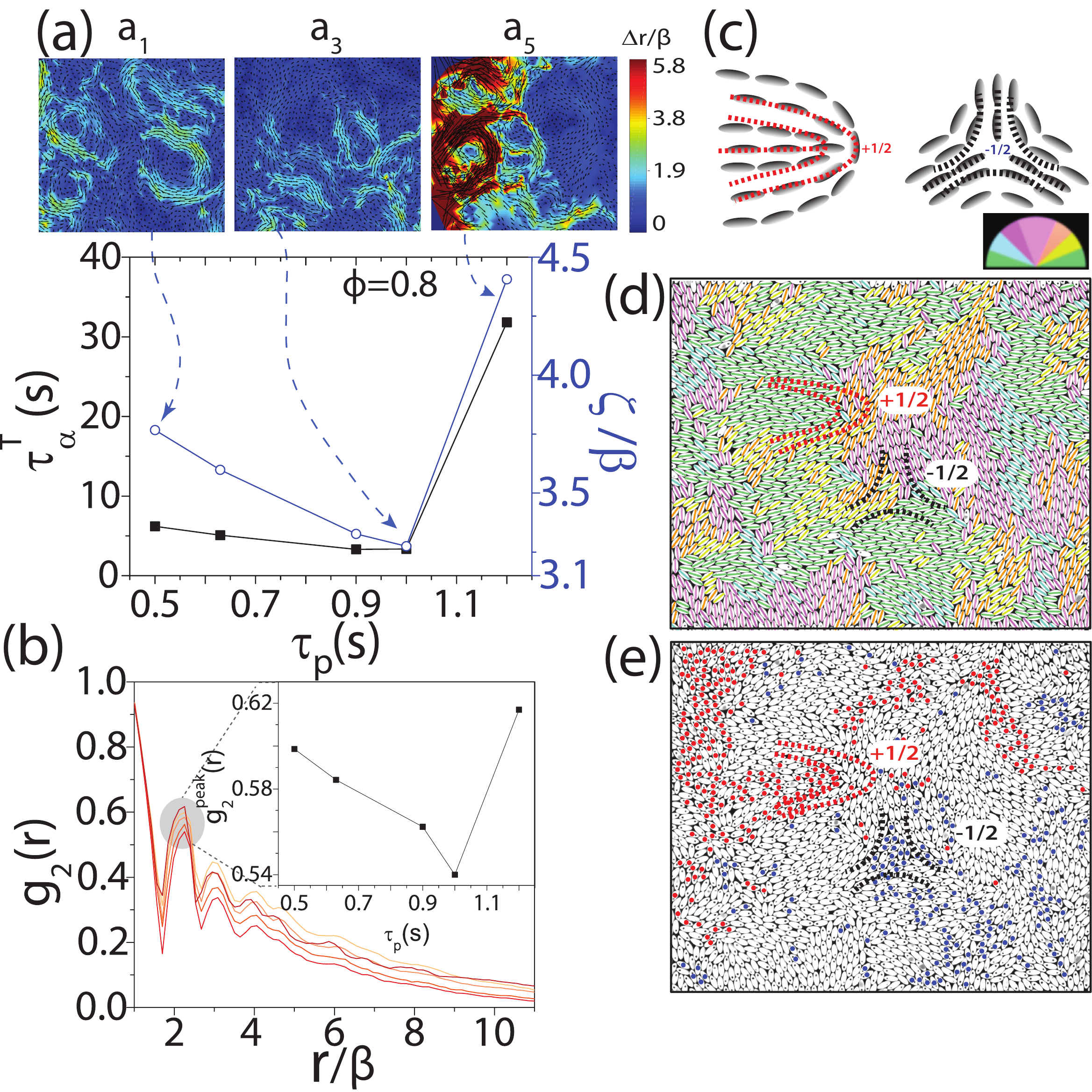}
\caption{(a)  Translational  relaxation time $\tau_{\alpha}^T$ (black squares) and the normalized swirl correlation length $\zeta/\beta$ (blue circles) vs $\tau_{p}$ for $\phi=0.8$.  Particle displacement maps over  $\tau_{\alpha}^T$ for $\phi = 0.8$ for $a_{1}$, $a_{3}$  and $a_{5}$ corresponding to the data points shown by the dashed curved arrows. Color bar shows the magnitude of scaled particle displacement $\Delta r/\beta$ over $\tau_{\alpha}^T $. (b) Static orientational correlation function $g_{2}(r)$ vs $r/ \beta$ for various $\tau_{p}$ .  Inset shows the height of the first peak of $g_{2}(r)$ as a function of $\tau_{p}$. (c) Schematic of   topological disclination defects of $\pm$1/2 strength, respectively.  (d) Orientational order field maps at $\phi=0.8$.  The red and black curved dashed lines depict the $+$1/2 and $-$1/2 disclination sites. We identified these defects by color-coding particle orientations as per the color-wheel. (e) Top 20 $\%$  translationally most-mobile (red filled circles) and least-mobile (blue filled particles) at $\phi=0.8$. 
} 
\label{Figure3}
\end{figure}

At deeper supercooling ($\phi>0.79$), however $\zeta$ grew substantially and the swirls interfered with glassy slowing down. The top panels in Fig. \ref{Figure3}(a) show the representative particle displacement maps, calculated over $\tau_{\alpha}^T$ and color-coded according to their magnitude for $\phi=0.8$ (Movie S5). In dense liquids of passive ellipsoids with $p= 2.5$, the dynamics are essentially frozen at this $\phi$ \cite{mishra2013two}. Strikingly, $\zeta$ showed a clear non-monotonicity with activity, mirroring the behaviour of $\tau_{\alpha}^T$ (Fig. \ref{Figure3}(a) bottom panel).  To glean insights into the re-entrant behaviour, we calculated the static orientational correlation function $g_2(r)$ (Supplemental Material S9) at $\phi=0.8$ (Fig. \ref{Figure3}(b)). At all $r$, $g_2(r)$ also showed clear re-entrant behaviour with $\tau_p$ (see inset to Fig. \ref{Figure3}(b)). For small and large values of $\tau_p$, steric repulsion, and strong activity-mediate attraction, respectively, promote nematic alignment, while at intermediate activities, orientational disorder in the void spaces dominates and suppresses overall nematicity.

The nematic ordering of the swirls naturally led us to look for topological disclination defects of $\pm1/2$ strength \cite{marchetti2013hydrodynamics,giomi2014defect}. In active nematics, even sans momentum conservation \cite{grossmann2020particle, shi2013topological}, like our system, curvature-driven stresses render $+1/2$ defects motile and $-1/2$ ones immotile \cite{bacteria}. The fingerprint of these defects in the orientational order field is shown in Fig. \ref{Figure3}(c) and  Fig. \ref{Figure3}(d). As expected, only the $+1/2$ defects were motile, and this was the case for all the activities  (See Supplementary Material S12, Fig. S13 ). We however do not see measurable local density differences around $+1/2$ and $+1/2$ defect cores (See Supplementary Material S13, Fig. S14 and Table S2). More importantly, the top 20\% translationally most-mobile (least-mobile) particles over the cage-breaking time $t^*$, obtained from the peak of the non-Gaussian parameter (Supplemental Material S14 and Fig. S15), were in the vicinity of $+1/2$ ($-1/2$) defects (Fig. \ref{Figure3}(e) and Movie S6). Thus, these mobile topological defects act as conduits for structural relaxation and preclude the system from undergoing dynamical arrest at even high $\phi$.  

\begin{figure}[tbp]
\includegraphics[width=0.6\textwidth]{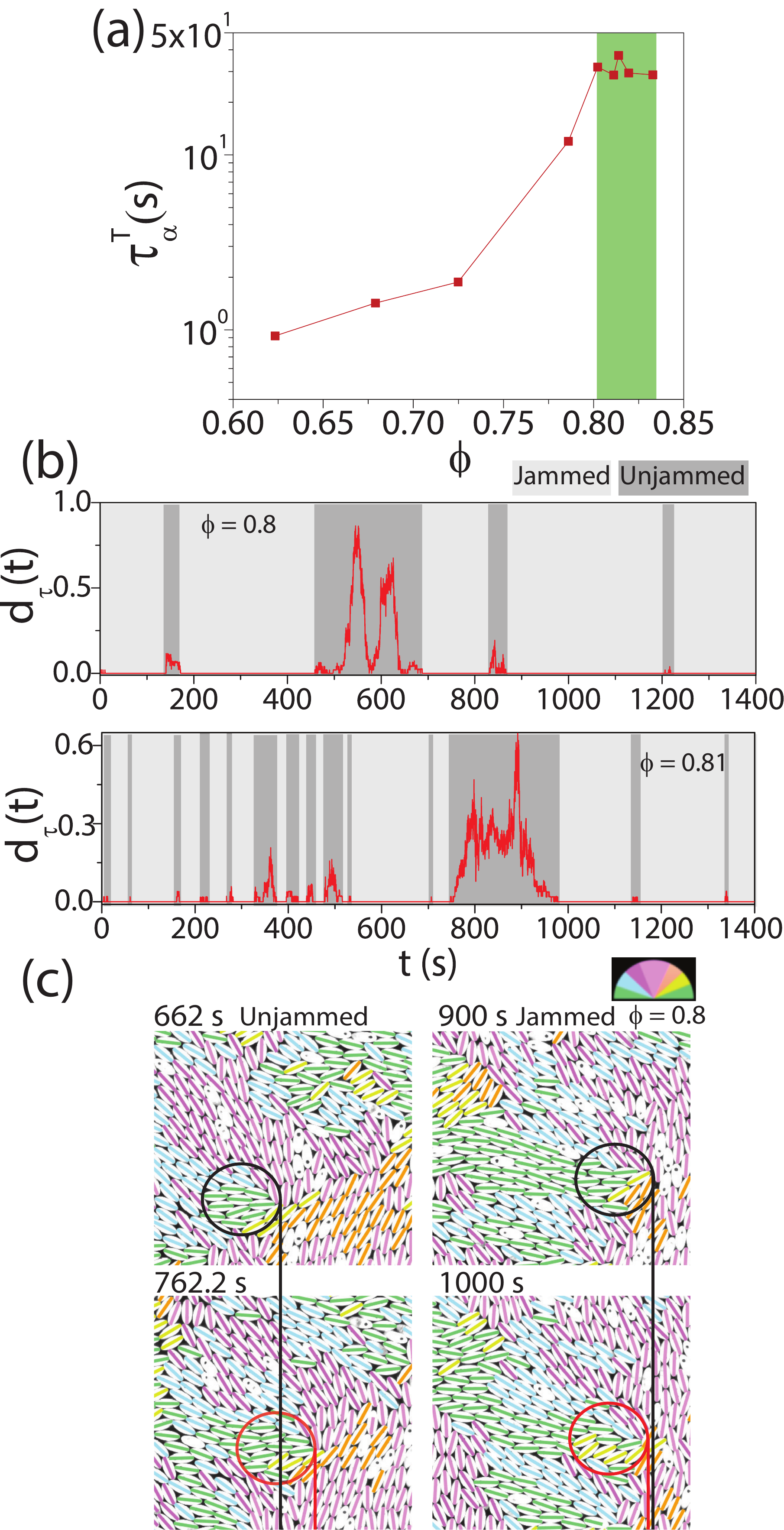}
\caption{(a)  Translational  relaxation  times  $\tau_{\alpha}^T$ vs $\phi$ for $a_{5}$. Note the saturation in values of  $\tau_{\alpha}^T$ at large $\phi$ (green shaded region). (b) Displacement overlap function $d_{\tau}(t)$  for $a_{5}$ at $\phi=0.80$ and $\phi=0.81$. In the intermittent phase we see bursts of activity in the system, a characteristic of jamming-unjamming. (c) Orientational order field maps at $\phi=0.80$.  Dynamics of +1/2 defects in the jammed and unjammed regions over a duration of 100 $s$, respectively. } 
\label{Figure4}
\end{figure}

The influence of these mobile $+1/2$ defects on structural relaxation was most striking for $a_5$ since the active curvature stresses are largest here - the strong activity-mediated attraction also promotes strong nematic ordering of the ellipsoids. Here, we observed that both $\tau_{\alpha}^T$ (Fig. \ref{Figure4}(a)) and $\tau_{\alpha}^R$ (Fig. S16) in fact saturated for $\phi>0.8$. Furthermore, since our particles are rough, we observed the build-up and release of the active stresses manifest as jamming-unjamming events (Movie S7) \cite{peshkov2016active}. We quantified these events through the coarse-grained displacement overlap function $d_{\tau}(t)$. To calculate this function, we binned the field of view into 30 square boxes of side $4.6\alpha$ and then for each box and over $\tau_{\alpha}^T$, we calculated the average displacement, $\bar{d}_{\tau}(t)$, of the particles within the box.  $d_{\tau}(t)=1$ if $\bar{d}_{\tau}(t)>\alpha$ the particle size) and zero otherwise. Figure \ref{Figure4}(b),  shows these jamming-unjamming events for $\phi = 0.8$ (top panel) and $\phi=0.81$ (bottom panel). Figure \ref{Figure4}(c) shows the dynamics of  $+1/2$ defects in the jammed and unjammed regions. We found that the $+1/2$ defects were mobile in the unjammed region and almost stationary in the jammed region. Thus, these observations bring to the fore the crucial role of defects in glassy slowing down.

Collectively, our study  shows that understanding glassy slowing down in generic dense active matter needs incorporating the rich physics of motile topological defects 
\cite{ doostmohammadi2018active, saw2017topological, kawaguchi2017topological, giomi2014defect, shi2013topological}, unique to active liquids, with the well-developed concepts for classical liquids with sticky attraction \cite{pham2002multiple, eckert2002re, mishra2013two, ginot2015nonequilibrium, klongvessa2019active}. A crucial experimental advance made here that allowed this inference was our strategy to tune  the persistence time of the active particles. While at moderate supercooling, liquids of active ellipsoids behave like a `hot' sticky classical liquid in both translational and orientational DoF, at deeper supercooling, this correspondence breaks  even as motile $+1/2$ disclination defects inhibit glassy slowing down by providing a new pathway for structural relaxation. Remarkably, at the highest activity, our experiments revealed intermittent jamming-unjamming of these $+1/2$ disclinations. Although our experiments were restricted to particles with run-and-tumble dynamics, our approach to tune activity is far more generic. It should now be possible to explore the role of both particle shape and the nature of activity on active glass dynamics \cite{arora2021emergent}. The future looks exciting now that well-controlled experiments with synthetic active matter can keep abreast with simulations \cite{bi2016motility}.\\

We thank I. Pagonabarraga, S. Nandi, P. Chaudhuri, M. Rao, S. Ramaswamy and C. Dasgupta for useful discussions. The research was funded by Department of Science and Technology (DST), Govt. of India through a SwarnaJayanthi fellowship grant(2016–2021) to RG. PA and RG designed experiments and wrote the paper. PA performed experiments and carried out data analysis. AKS contributed to project development. 


\newpage

\end{document}